\documentclass[a4paper,11pt]{article}
\usepackage{jinstpub} 
\usepackage{lineno}
\usepackage{algorithm}
\usepackage{algpseudocode}
\usepackage{enumerate}
\usepackage{adjustbox}
\usepackage{wrapfig}
\usepackage{subcaption}

\title{
Parallel CPU- and GPU-based connected component algorithms for event building for hybrid pixel detectors}

\author[a, b]{T. Čelko,}
\author[a]{F. Mráz,}
\author[b]{B. Bergmann}
\author[b, c]{and P. Mánek}
\affiliation[a]{Department of Computer Science Education, Charles University,\\
Malostranské nám. 25, 118 00 Prague 1, Czech Republic}
\affiliation[b]{Institute of Experimental and Applied Physics, Czech Technical University\\
Husova 240/5, 110 00 Prague 1, Czech Republic}
\affiliation[c]{Department of the Physics and Astronomy, University College London,\\
Gower Street, London WC1E 6BT, UK}

\emailAdd{celko@ksvi.mff.cuni.cz}

\abstract{

The latest generation of Timepix series hybrid pixel detectors enhance particle tracking with high spatial and temporal resolution. However, their high hit-rate capability poses challenges for data processing, particularly in multidetector configurations or systems like Timepix4. Storing and processing each hit offline is inefficient for such high data throughput. To efficiently group partly unsorted pixel hits into clusters for particle event characterization, we explore parallel approaches for online clustering to enable real-time data reduction. Although using multiple CPU cores improved throughput, scaling linearly with the number of cores, load-balancing issues between processing and I/O led to occasional data loss.
We propose a parallel connected component labeling algorithm using a union-find structure with path compression optimized for zero-suppression data encoding. Our GPU implementation achieved a throughput of up to 300 million hits per second, providing a two-order-of-magnitude speedup over compared CPU-based methods while also freeing CPU resources for I/O handling and reducing the data loss.}


\begin{document}
\maketitle
\flushbottom

\section{Introduction}
\label{sec:intro}

The Timepix family of pixelated semiconductor detectors \cite{Sriskaran_2024}
represents a significant advancement in particle detection technology, offering enhanced spatial and temporal resolution. With each new generation of detectors, the performance has progressively improved.

A particularly notable development within this family is the Timepix3 detector \cite{timepix3_poikela_2014}, which introduced a data-driven mode that allows precise measurements in high-flux radiation environments. The structure of the data produced by this detector can vary depending on the chosen operating mode. Typically, a \emph{pixel hit} is registered for each pixel that records a non-zero induced voltage. Each pixel hit consists of the following information. The \textit{x} and \textit{y} coordinates indicate the \textit{spatial position} of the hit pixel. Then, there is the timestamp corresponding to the moment when the voltage exceeds a predefined threshold, referred to as the \textit{time of arrival} (ToA). Lastly, the \emph{time over threshold} (ToT) indicates the magnitude of the induced voltage.

Notably, data rates can surpass 40 million hits per second in high-flux scenarios, necessitating highly efficient data processing techniques. 
A significant bottleneck in current data processing workflows arises from the need to organize individual pixel hits into meaningful events based on their temporal and spatial proximity, a process referred to as clustering. 
A cluster is the detector's response to ionizing radiation interacting with the sensor. The radiation leaves traces whose shape and energy content indicate -- to some extent -- particle type, particle energy, or the type of interaction. 
Informally, a cluster is the trace left by a particle or several coincident particles that participated in an event, e.g., a decay of a radioactive ion. 
This paper demonstrates how to accelerate clustering through parallel computing on both CPU and GPU architectures. Fast clustering has numerous applications, some of which are outlined below. First, computing cluster features that can be used for on-the-fly data filtering and data compression, reducing the storage space requirements.
Second, clustering is a prerequisite for real-time particle identification and classification.
Lastly, clustering can improve contrast in imaging applications by correcting for the charge-sharing effect~\cite{xray_jakubek}.

Although the presented results were obtained using Timepix3, they are transferable to other detectors functioning in either data-driven or frame-based mode, including the higher-performing Timepix4, where the requirement for processing efficiency is even more pronounced.

 \section{Preliminaries -- clustering}
 \label{sec:clustering}

Clustering is the process of grouping temporally and spatially coincident hits together. Formalizing this definition can result in multiple application-specific definitions, and we present three variants of such a definition. 

Given a sequence of hits $H = \{h_1, h_2, \dots h_n\}$, we say that subset $C = \{h_{i_1},h_{i_2}, \dots h_{i_m}\}$ is a \textit{cluster} in $H$ if all of the following conditions are satisfied:

\begin{enumerate}[(i)]
    
\item There is no hit $h\in H \setminus C$ such that $C' = C \cup \{h\}$ is a cluster (the \textit{maximality} condition). 
\item  For all $h_a, h_b \in C$, there exists a sequence $P = p_1, p_2, \dots p_l$, $P \subseteq C$ such that $p_1 = h_a$, $p_l = h_b$, and for all $i \in \{1,2,\dots l-1\}$, hits $p_i$ and $p_{i+1}$ are spatially 8-neighboring (the \textit{spatial neighborhood} condition).
 \item Based on the variant of the definition, one of the following conditions holds
\begin{enumerate} 

  \item For all sequences $P$ as defined in (ii), it also holds $|\mathrm{toa}(p_i) - \mathrm{toa}(p_{i+1})| \leq \Delta t_{\mathrm{max}}$ for all  $i \in \{1,2,\dots l-1\}$,  where $\mathrm{toa}(h)$ represents the time of arrival of hit $h$  (\textit{dynamic local-time-neighborhood} variant used in~\cite{clustering_celko_2023})
  \item For all $h \in C$, either $\mathrm{toa}(h) = \min\{\mathrm{toa}(h)|h\in C\}$ or there exists $h' \in C \setminus \{h\}$ with $\mathrm{toa}(h) - \Delta t_{max} \leq \mathrm{toa}(h') \leq \mathrm{toa}(h)$  (\textit{dynamic global-time-neighborhood} variant used in~\cite{system_manek2018,tracklab_manek}).
 \item  For all $ h_a, h_b \in C$, $|\mathrm{toa}(h_a) - \mathrm{toa}(h_b)| \leq \Delta t_{\mathrm{max}}$ (\textit{static time-neighborhood} variant used in~\cite{detecting_meduna_2019})
 \end{enumerate}
\end{enumerate}

For large values of $\Delta t_{\text{max}}$ or in the case of small clusters, these definitions yield identical results. However, as cluster size increases, particularly when $\Delta t_{\text{max}}$ is less than the expected maximum duration of the cluster (time span), the differences between the approaches become more pronounced. In the following, unless otherwise specified, the variant with the dynamic local neighborhood (condition (iii)(a)) will be used for most applications.

In the following subsections, we outline the core concepts of the existing state-of-the-art clustering approaches.

\subsection{Frame-based clustering} 
\label{sec:frame based clustering}
For detectors operating in a shutter-based mode, such as Timepix~\cite{timepix_llopart}, the output data can be interpreted as a sequence of image frames (similar to a standard camera). Each frame can be processed individually; we will treat it as a graph
$G=(V,E)$. In this graph, each detected hit in the frame is a node in $V$. If two hits are adjacent in space and time, they are connected with an edge in $E$. Clustering in this context can be reduced to finding connected components in the graph $G$, which can be efficiently solved using graph traversal algorithms (e.g., depth-first-search). However, the use of a shutter introduces a potential loss of information. For example, two spatially overlapping clusters in a single frame could be mistakenly merged into a single cluster, even if there was a considerable time difference between them. Conversely, a cluster occurring near the shutter's closing time could be incorrectly split into multiple frames.

 \subsection{Data-driven clustering}
In the case of a detector with a data-driven output mode (like Timepix3~\cite{timepix3_poikela_2014}), we can utilize an algorithm that processes data on a pixel-by-pixel basis, avoiding the loss of the information introduced by the shutter in the frame-based mode. 

Algorithm~\ref{alg:data-driven clustering} addresses three possible scenarios for processing a hit based on the set of neighboring clusters to which the hit might belong. If no adjacent cluster exists, a new cluster is created. If the hit can be added to a single cluster, it is incorporated into that cluster. However, if the hit can be part of multiple clusters, those clusters should additionally be merged. 

The key difference between existing approaches lies in how neighboring clusters are identified for each hit and which data structures they use, e.g., in the implementation of the \textit{findNeighborClusters} method. One approach is to iterate over the open clusters, conducting a hit membership query for each of them. Here, we need a data structure that allows fast membership queries such as a hash set (with expected constant-time query) or quadtree~\cite{quadtree_samet_1984} (with worst-case logarithmic time w.r.t. the sensor width). A different approach~\cite{system_manek2018} is to store references to currently open clusters for each pixel in the matrix, eliminating the need to iterate over all open clusters. In exchange, extra work in the form of pointer bookkeeping is required. As our baseline, we chose to refine the latter approach to fit the local-time-neighborhood cluster type.
\begin{algorithm}
\caption{High-level data-driven clustering algorithm \cite{system_manek2018}}
\label{alg:data-driven clustering}
\begin{algorithmic}[1]

\Procedure {$\text{processHit}$}{$\mathit{hit}$}
\State $\mathit{N} \gets \text{findNeighborClusters(}\mathit{hit}\text{)}$
\If{$|N| = 0$}
    $\text{createNewCluster(}\mathit{hit}\text{)}$
\ElsIf{$|N| = 1$}
    $\text{addHitToCluster(}\mathit{hit}, \mathit{N}\text{.first()}\text{)}$
\Else \State $\mathit{newCluster} \gets \text{ mergeClusters(}\mathit{N}\text{)}$  
   \State $\text{addHitToCluster(}\mathit{hit}, \mathit{newCluster}\text{)}$
\EndIf
\State $\text{closeAndDispatchOldClusters()}$
\EndProcedure
\end{algorithmic}
\end{algorithm}

\section{Methodology -- CPU-based parallelization}
This section describes different ways to distribute clustering work among multiple CPU threads.
\subsection{Step-based parallelization}
\label{sec:Step-based parallelization}
The step-based parallelization splits the clustering algorithm into independent stages that can be executed concurrently in a pipeline. Beyond potential speed improvements, this approach enhances the algorithm's modularity. The clustering process is divided into the following steps:

 \paragraph{Input Reading:} This step reads the hits from an input file or detector readout, preparing hits for further processing.
    
\paragraph{Hit Calibration:} Raw hits from Katherine's readout are converted into a format that includes energy information, which is more suitable for analysis. Additionally, time values (arrival time and time over threshold) are converted from clock ticks to nanoseconds.
    
\paragraph{Time Sorting:} Hits from the Timepix3 readout are not sorted w.r.t. time. However, they are $t$-ordered for some positive constant t. A sequence of hits $h_1,h_2,\ldots,h_l$ is $t$-ordered if for all indices
$1\le i < j \le l$, it holds $\mathrm{toa}(h_i) < \mathrm{toa}(h_j) + t$ (see partial orderliness in~\cite{clustering_celko_2023}).
Hence, additional sorting is required to achieve a fully time-ordered sequence of hits. For that, we utilize a priority queue. Skipping sorting could complicate clustering, as the difference between $t$ and $t_c$ (the expected cluster time span) would require retaining the clusters in memory for a longer period, increasing the number of simultaneously active clusters. In the implementation, this step is merged with the hit calibration step for performance reasons. 
    
\paragraph{Clustering:} The sorted hits are grouped into clusters, utilizing Algorithm~\ref{alg:data-driven clustering}.
    
\paragraph{Cluster Outputting:} Clusters are written to output files. This step may involve filtering or analyzing clusters to reduce the data volume.

\subsection{Data-based parallelization}
\label{sec:data based parallelization}
Data-based parallelization is a technique that is orthogonal to step-based parallelization. Instead of dividing the algorithm into sequential steps, data-based parallelization involves partitioning the data into blocks and assigning each block to a computing worker. The primary consideration is how to partition the data among workers to minimize the work required to merge the data streams of potentially incomplete clusters.
\subsubsection{Hit count splitting}

The most straightforward method of partitioning the data is to divide the data stream into consecutive blocks of fixed size and distribute them among workers. The advantage of this approach is that it ensures a balanced distribution of workload across workers. However, to estimate the expected number of split clusters, even with time-sorted hits, we would expect at least $\frac{\textit{clusterSize} - 1}{\textit{clusterSize}}$ clusters to be split per block. To identify the split clusters, all hits within the $t$-unsortedness region around the block borders must be checked (because each block is sorted individually), as they may belong to a split cluster. 

\subsubsection{Spatial splitting}
An alternative approach is to distribute the hits based on their spatial coordinates (e.g., dividing the sensor into quadrants). However, this method risks unbalanced workload distribution if the data is not uniformly spread across the sensor (e.g., in the case of a narrow beam of particles). Furthermore, it does not scale well with an increasing number of regions, as smaller regions increase the likelihood of splitting clusters. For example, with a mean cluster width of 4 pixels and four quadrants, we could still expect approximately $p_{\text{border}} = 4\%$ of clusters to lie on the borders, leading to around 2\% of clusters being split. To achieve a comparable number of border pixels with hit-based splitting, a very large block size would be required, corresponding to $\textit{maximumHitrate} \cdot \frac{t}{p_{\text{border}}}$ hits, which evaluates to $6 \times 10^{5}$ hits per window for $t = 600\,\text{$ \mu$s}$ and $\textit{maximumHitrate} = 40\,\text{MHit/s}$.
\subsubsection{Temporal splitting}
\label{sec:temporal data splitting}
Lastly, we can partition the hits based on their timestamps. With a sufficiently small time window, this method is likely to ensure a balanced workload distribution. Additionally, it scales effectively with the number of cores, as time windows can be assigned to workers in a round-robin fashion. Since hits are temporally contiguous, we only need to examine the $\Delta t_{\text{max}}$-time neighborhood around each border (typically $\Delta t_{\text{max}} \ll t$). For example, for a time window of $\textit{splitWindowSize} = 10000\,\text{ ns}$ (which is still relatively small), with a temporal hit neighborhood of $\Delta t_{\text{max}} = 200\,\text{ ns}$ and an expected cluster timespan of $\Delta t_{\text{cluster}} = 25\,\text{ ns}$, this yields an expected fraction of border clusters around 2\%, with an expected number of split clusters below 1\%. For larger clusters, the $\textit{splitWindowSize}$ can be increased to maintain similar performance.      

\subsection{Merging split clusters}
While data-based parallelization (Section~\ref{sec:data based parallelization}) offers the possibility to utilize even more CPU cores, it also introduces new problems with incorrectly split clusters -- similar to the shutter-induced problem in frame-based mode (Section~\ref{sec:frame based clustering}).
For the rest of the section, we consider using temporal data splitting (Section~\ref{sec:temporal data splitting}). We will approach this as follows. If the cluster is sufficiently far from the split time window border, we know that it is correct and does not need to undergo a merging process. Otherwise, we call it a $\textit{border cluster}$. Additionally, we keep track of all clusters that are temporally $\Delta t_{\text{max}}$-close to the ``latest time'' given by the first hit ToA of the latest cluster and call these $\textit{open clusters}$. 

To process the current border cluster, we find all open clusters that can merge with it. Note that multiple clusters could be merged in this process. During the merging, we combine all found clusters into a single cluster, storing it in the open clusters at the position of the cluster with the smallest first time of arrival among the merged clusters. In this way, the clusters are kept sorted in time. Subsequently, as a performance optimization, we invalidate all other merged clusters instead of eagerly removing them. Invalid clusters are periodically removed when closing ``old'' open clusters.
Note that in open clusters, non-border clusters can also be present (e.g., in the case when we have a border cluster spanning $T = (t_\text{start}, t_\text{end})$ time interval and non-border cluster spanning $T' \subset T$).

Throughout this process, we have not yet addressed the method for efficiently determining whether two clusters are mergeable. Our approach to identifying mergeable clusters prioritizes the fast elimination of clusters that cannot be merged. To achieve this, we employ a cascading strategy consisting of three steps, each with progressively increasing temporal complexity.

\begin{figure}[H]
\centering
\includegraphics[width = 0.75 \linewidth]{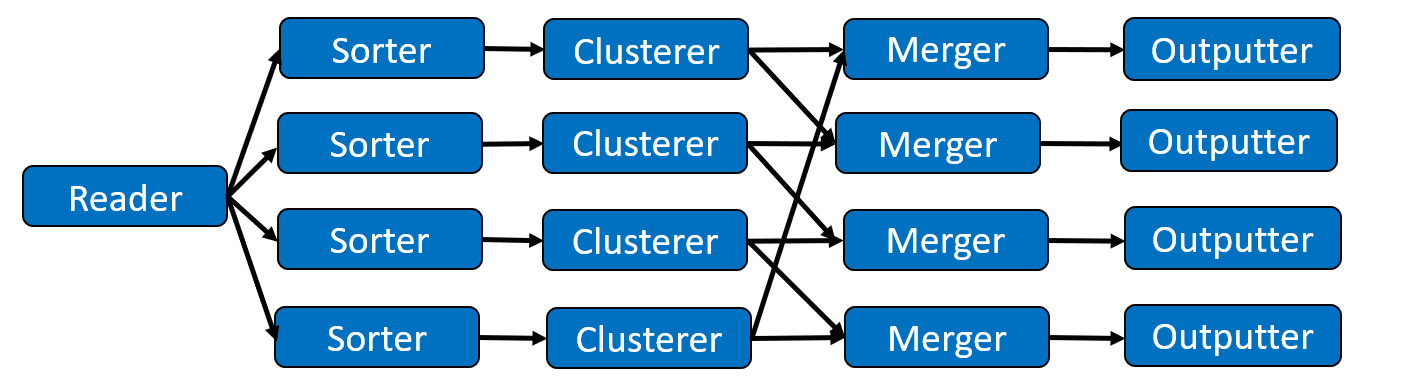}
\caption{Clustering computation graph combining both step-based (Section~\ref{sec:Step-based parallelization}) and data-based (Section~\ref{sec:data based parallelization}) parallelizations with $n_\mathrm{datalines}=4$ data lanes. Arrows denote the data flow, and each box represents an independent worker thread. \label{fig:CPU clustering pipeline}}
\end{figure}

\begin{enumerate} \item If the temporal distance between two clusters exceeds the maximum threshold $\Delta t_{\text{max}}$, they cannot be merged. The temporal distance $\mathit{dt}_{c_1, c_2}$ between clusters $c_1$ and $c_2$ is defined as  \(\mathit{dt}_{c_1, c_2} = \max\{\text{startTime(}c_1\text{)} - \text{endTime(}c_2\text{)}, \text{startTime(}c_2\text{)} - \text{endTime(}c_1\text{)}, 0\}\).
This step can be executed in constant time using the precomputed start and end times of each cluster.
\item If the bounding boxes of the clusters do not intersect, the clusters are not spatially neighboring and cannot be merged. The bounding box, defined as the smallest rectangle (aligned with the sensor) enclosing the cluster, is computed once upon cluster creation. Although the initial bounding box computation requires linear time $O(n)$ (with respect to cluster size), checking for intersections between bounding boxes is performed in constant time (independent of the size of the cluster) by simply comparing the coordinates of the bounding box vertices.

\item If neither of the previous conditions excludes a merge, a full merge check is performed. A naive pairwise comparison of pixels would require $O(n_1 \cdot n_2)$ time, where $n_1$ and $n_2$ are the pixel counts of the clusters. Instead, we store 
pixels of the larger cluster that are in the intersection of the bounding boxes
in a 2D array. We then check if any pixel from the smaller cluster has a spatial neighbor in the array. The larger cluster is chosen for the 2D array due to the higher computational cost of neighbor checking compared to array insertion. Assuming $n_1 > n_2$, the neighbor check requires $O(9n_1)$ time (due to 9 spatial neighbors), while inserting and removing a cluster from the 2D array takes $O(2n_2)$. The overall time complexity is $O(n_1 + n_2)$. 

\end{enumerate}
 However, with increasing the degree of parallelization (increasing $n_\mathrm{datalanes}$; see Figure~\ref{fig:CPU clustering pipeline}), single-threaded merging does not suffice. Fortunately, it can be parallelized, assuming the clusters do not span more than half of the time window. The idea is that each merging worker will be responsible for merging clusters at a particular border time, distributed in a round-robin fashion. For the $i$-th time window, we send the clusters from the lower half of the window to the $(i \mod n_\mathrm{datalanes})$-th merging worker, and those from the upper half are sent to the $((i+1) \mod n_\mathrm{datalanes})$-th merging worker, as illustrated by Figure~\ref{fig:CPU clustering pipeline}. The parallel data streams may or may not be concatenated in the next step, depending on the use case.

\section{Methodology - GPU-based clustering}
Although we have not encountered any prior use of GPUs for clustering data from Medipix pixel detectors in the literature, the problem of labeling connected components in an image is well-established in the field of image processing \cite{grid_clustering1_allegretti,grid_clustering2_playne,GPU_CCL_KALENTEV}. However, instead of reducing our task to a conventional 2D connected component labeling (CCL) problem, we have developed a specialized approach tailored to the characteristics of our data. The following considerations drove this decision:

\begin{itemize} \item \textbf{Zero Suppression.} Timepix3 data is received as a stream of hit pixels, with zero pixels being suppressed. This feature can be leveraged to simplify the problem. For example, in the case of a linear diagonal track consisting of $n$ hit pixels, we only need to process $O(n)$ pixels. In contrast, conventional matrix-based methods require processing $O(n^2)$ pixels in the worst-case scenario. By focusing on non-zero data points, our approach reduces computational complexity.
\item \textbf{Data-Driven Mode.} On the other hand, the data-driven nature of the detector introduces an additional challenge. Transforming into a matrix-based CCL problem would require dividing the hits into frames, which inherently involves partially solving the clustering problem. Moreover, straightforward frame partitioning by closing a frame after a maximum time interval $\Delta t_{\text{max}}$ without new hits could lead to multiple hits sharing the same spatial coordinates within the frame, potentially resulting in incorrect overlapping clusters. (For further discussion, refer to the experiment evaluating the correctness of temporal clustering in ~\cite{clustering_celko_2023}).

\end{itemize}

 \subsection{Our zero-suppression based algorithm}

Our proposed parallel algorithm follows the same high-level logic as the Algorithm~\ref{alg:data-driven clustering}, with modifications in the data representation aimed at minimizing GPU memory access. The detailed steps of the algorithm are outlined in Algorithm~\ref{alg:High level GPU clustering}. 

Using a straightforward implementation in Step 1 of Algorithm~\ref{alg:High level GPU clustering} would require checking the $t$-border of each buffer (typically by the CPU) to identify clusters that span multiple buffers. However, we optimize this by applying Algorithm~\ref{alg:GPU buffer filling}, which reduces the border checking to only the $\Delta t_{\text{max}}$ region, as the algorithm completely eliminates cluster splitting caused by unsortedness. The main idea is to keep an empty space at the end of the buffer to handle delayed hits. The constant $t_\text{closing}$ is set to further decrease the splitting probability, as with high probability, the recorded value $\mathit{toa}_\mathrm{max}$ would represent a cut within a cluster.

\begin{algorithm}
\caption{High-Level GPU clustering algorithm \label{alg:High level GPU clustering}}
\begin{algorithmic}

\State \textbf{Step 1:} Populate the data buffer by applying Algorithm~\ref{alg:GPU buffer filling} to reduce border checking and mark it as n-use until the clustered data is retrieved from GPU.

\State \textbf{Step 2:} Transfer hits from host to device.

\State \textbf{Step 3:} Sort hits by their time of arrival. A fast option is the parallel radix sort algorithm~\cite{radix_chapter_bell}.

\State \textbf{Step 4:} Partition hits into chunks, and cluster hits within each chunk in parallel. Each thread is provided with an auxiliary 256 × 256 matrix that stores references to the last hit occurring at each pixel, enabling efficient neighbor checking.

\State \textbf{Step 5:} Use the algorithm from the previous step to group clusters at the borders of adjacent chunks. Specifically, the $i$-th thread handles the border hits at the start $(i+1)$-th chunk, using the state of the $i$-th auxiliary matrix.

\State \textbf{Step 6:} Sort clusters by their minimum time of arrival, causing the hits from the same cluster to form adjacent memory blocks.

\State \textbf{Step 7:} Restore the state of auxiliary matrices used for clustering by iterating over all hits within the chunk rather than processing the entire matrix.
\State \textbf{Step 8:} Transfer hits and labels from device to host.

\State \textbf{Step 9:} Send transferred data for subsequent processing tasks and mark the buffer memory as reusable.
\end{algorithmic}
\end{algorithm}

\begin{algorithm}
\caption{Hit buffer filling}
\label{alg:GPU buffer filling}
\begin{algorithmic}[1]
\State $\text{constant } b $
\Comment{Maximum buffer size, typically in the order of $10^7$ or more. }

\State $\text{constant } t_\text{closing}$
\hfill\Comment{\parbox[t]{.7\linewidth}{A small time interval used to compute the final boundary time,  decreasing the probability of cluster splitting. It is set to the maximum expected cluster duration.}}

\State $\text{constant } b_t$ \Comment{Maximum expected number of hits that can arrive within $t + t_\text{closing}$ time.} 

\State $\text{global variable } \mathit{toa}_\text{max} \gets 0$
\Procedure {$\text{storeHit}$}{$\mathit{hit}, \mathit{buffer}, \mathit{nextBuffer}$}
\If {$\text{size(}\mathit{buffer}\text{)} < b - b_t$)}
    \State $\mathit{buffer} \gets \mathit{buffer} \cup \{hit\}$
    \State $\mathit{toa}_\text{max} \gets \text{max\{} \mathit{toa}_\text{max}, \text{toa(}\mathit{hit}\text{)} \text{\}}$
\ElsIf{$\text{toa(}\mathit{hit}\text{)} < \mathit{toa}_\text{max} + t_\mathit{closing}$}
    \State $\mathit{buffer} \gets \mathit{buffer} \cup \{hit\}$

\Else
    \State $\mathit{nextBuffer} \gets \mathit{nextBuffer} \cup \{hit\}$
\EndIf
\If{$\mathrm{toa(\mathit{hit})} - \mathit{toa}_\mathrm{max} > t + t_\text{closing}$}
\State $\text{sendToDevice(}\mathit{buffer}\text{)}$
\State $\mathit{buffer} \gets \mathit{nextBuffer}$
\EndIf
\EndProcedure
\end{algorithmic}
\end{algorithm}

We now explain the hit grouping process outlined in Step 4 of Algorithm~\ref{alg:High level GPU clustering}. For each hit \( h \), we store a reference to another hit within the same cluster, referred to as the \textit{parent} of \( h \). This forms a graph \( G = (\mathit{hits}, \{(h, \mathit{parent}(h)) \mid h \in \mathit{hits}\}) \), which represents an oriented forest, also known as a union-find data structure. This data structure facilitates fast \textit{union} and \textit{find} operations.

In the context of Algorithm~\ref{alg:data-driven clustering}, the function findNeighboringClusters(\textit{hit}) is implemented by invoking the \textit{find} operation for each of the 8 neighboring pixels plus the pixel itself. Conversely, addHitToCluster(\textit{hit, cluster}) and mergeClusters(\textit{clusters}) are implemented using the \textit{union} operation.

To ensure clusters are eventually sorted by arrival time and to facilitate quick access to the first pixel of a cluster, we maintain a \textit{time-invariant} property: \(\mathrm{toa}(h) \geq \mathrm{toa}(\mathrm{parent}(h))\).

\begin{figure}[H]
\centering
\includegraphics[width = \linewidth]{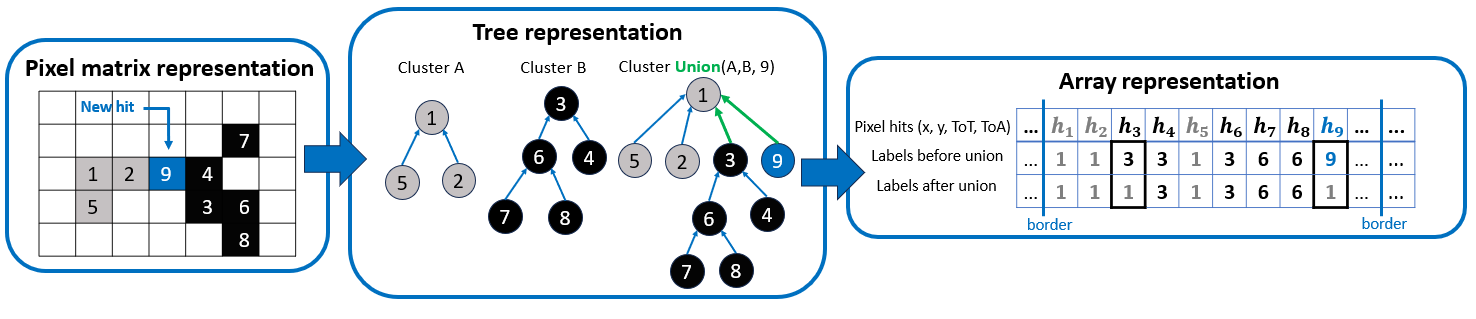}
\caption{Three different views of the hit data. Pixel matrix representation describes hits as pixels labeled by their index in the time-sorted buffer. The same hits can be represented by trees, showing how two trees are joined. Such a tree can be trivially implemented in a static array using tree parent pointers (labels). \label{fig:DFU representation}}
\end{figure}

Consequently, to preserve this time-invariant when performing the \textit{union} operation, we select the root with the smaller time of arrival as the new root, see Figure~\ref{fig:DFU representation}. For further optimization, \textit{path compression} is employed during each \textit{find} query, setting the parent of each node along the path to the root.

\section{Application to experimental data}

\begin{figure}[htbp]
    \centering
    \begin{subfigure}[b]{0.46\textwidth}
        \centering
        \scriptsize

\begin{tabular}{|p{3cm}|p{2cm}|p{2cm}|}
\hline
\textbf{Dataset} & \textbf{Mean cluster size} & \textbf{Standard deviation of cluster size} \\
\hline
$\gamma$, 59.6 keV from Am-241 & 2.46  & 2.15   \\
\hline
$\pi$, 40 GeV/c, 0°            & 7.22  & 27.35   \\
\hline
$\pi$, 40 GeV/c, 45°           & 23.33 & 33.47  \\
\hline
$\pi$, 40 GeV/c, 75°           & 60.27 & 64.33  \\
\hline
Pb, 385 GeV/c, 0°              & 131.84& 500.51\\
\hline
Pb, 385 GeV/c, 50°             & 26.51 & 274.99 \\
\hline
Pb, 385 GeV/c, 90°             & 20.45 & 356.57 \\
\hline
Pb, 385 GeV/c, 0°, subset      & 2231.45 & 363.69 \\
\hline
Pb, 385 GeV/c, 50°, subset     & 3622.54 & 860.56 \\
\hline
Pb, 385 GeV/c, 90°, subset     & 7258.26 & 5105.26 \\
\hline
\end{tabular}

        \caption{\label{fig:table dataset.png}}
    \end{subfigure}
    \hfill
    \begin{subfigure}[b]{0.43\textwidth}
        \centering
        \includegraphics[width=\textwidth]{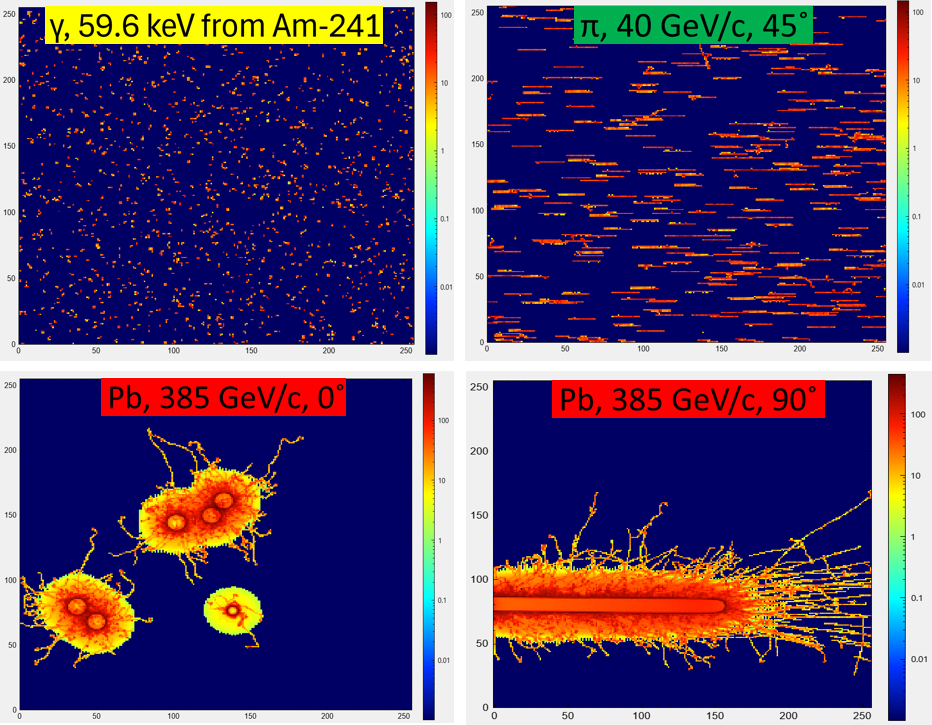}  \caption{\label{fig:dataset examples}}
    \end{subfigure}
    \caption{The benchmarking dataset cluster sizes in pixels (a) and the cluster examples from the particular datasets, the pixel color indicating the ToT (b). Pion and lead data were acquired during measurements at SPS CERN beam lines. The last three datasets were artificially created by selecting a subset of large clusters from the lead dataset.}
    \label{fig:cluster dataset major label}
\end{figure}

In this section, we evaluate algorithm efficiency across multiple real-world datasets, see Figure~\ref{fig:cluster dataset major label}. To accurately evaluate the clustering performance for variable-sized clusters, we developed datasets encompassing a range of cluster sizes, from small clusters consisting of only a few hits to larger clusters containing thousands of hits.

Let us first introduce how the benchmark is conducted. In the CPU clustering benchmark, the hits ($n_\text{hits} = 4,000,000$) are loaded into memory from a file, and the clock is started. The loaded hits are repeated with an offset in their time of arrival $n_\text{rep} = 25$ times and then clustered. The clock is stopped after processing the last hit. Each measurement is repeated 5 times.

For the GPU clustering benchmark, the data generation step is excluded, as GPU clustering throughput can exceed the data generation speed of the CPU. The clock starts once the host buffer is populated with data and ends after the clustered data is transferred back to the host.

To accurately assess throughput, we increase the data volume by configuring $n_\text{rep} = 250$. Additionally, we conduct a GPU warm-up and measure processing speed only from the second buffer onward to ensure that the initialization effects do not influence the performance metrics.

Both methods, independent of I/O speed, enable the measurement of clustering throughput beyond current readout capabilities.

The proposed parallel algorithms were also compared with state-of-the-art clustering~\cite{tracklab_manek} to verify their correctness. For most test datasets, we reached intersection over union~\cite{IOU_Rezatofighi_2019} (IoU) over 99.99\% with the exception of the large clusters (Pb) where the IoU was much lower due to varying cluster definition; see local vs. global neighborhood (variants (iii)(a) and (iii)(b) in Section~\ref{sec:clustering})%
. We were able to compensate for that and achieve an exact match by increasing  $\Delta t_\text{max}$ from 200 ns to 600 ns.

\subsection{Benchmarking results}
\begin{figure}[htbp]
\centering
        \centering
        \includegraphics[width=0.8\textwidth]{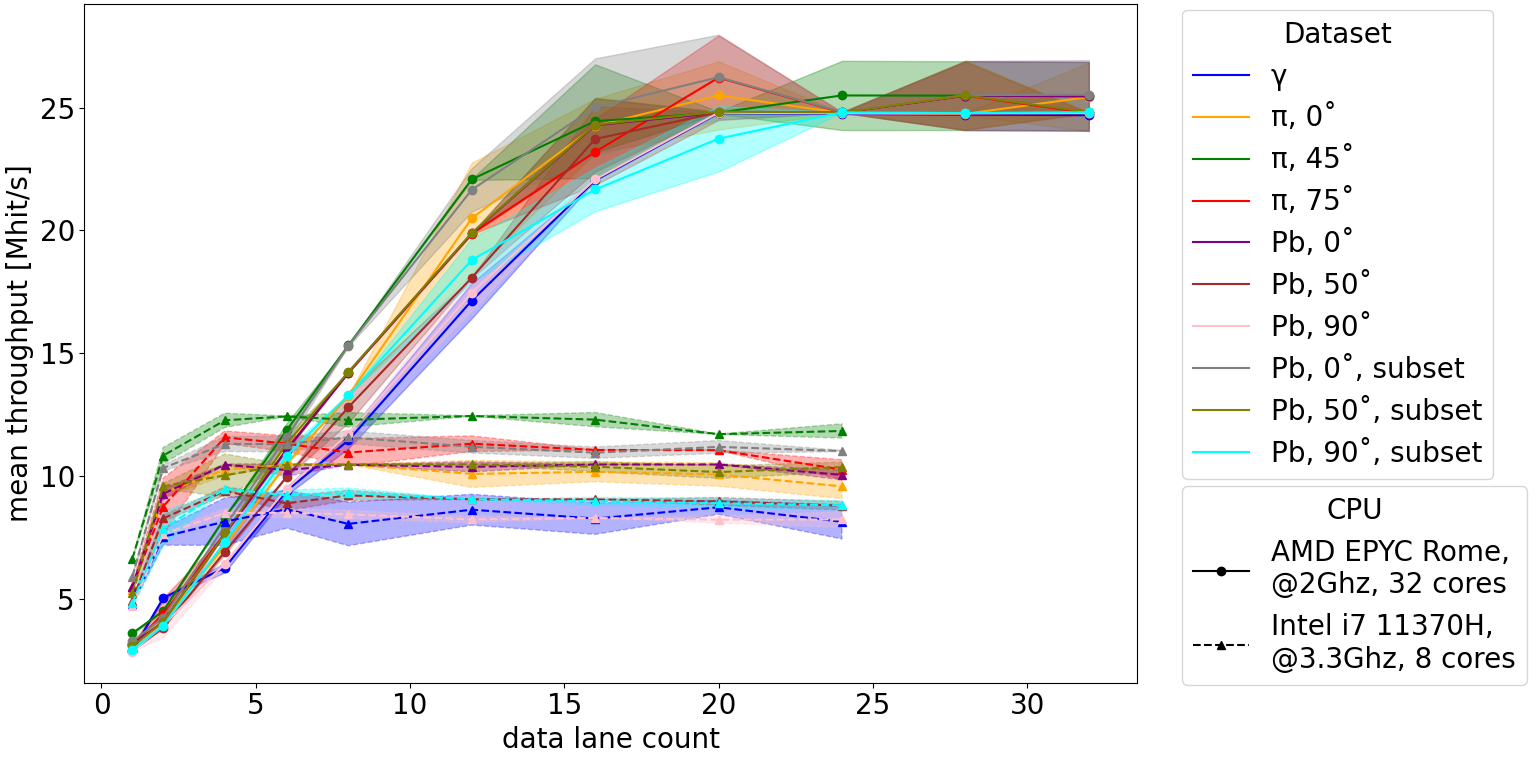}
        \caption{Dependence of parallel CPU clustering throughput on the number of data lanes (pipelines) in the computation graph (Figure~\ref{fig:CPU clustering pipeline}). One can see the throughput scaling with an increasing degree of parallelization. The difference in scaling among different CPUs is probably caused mainly by their number of cores and clock frequency. \label{fig:CPU benchmarking results}}
    \end{figure}

The measured clustering throughputs can be seen in Figure~\ref{fig:CPU benchmarking results}. The main purpose of these tests was to verify how the implementation scales with the degree of parallelization (a single data lane consists of calibrating plus sorting, clustering, and merging thread).    
    \begin{figure}[htbp]
    \centering
        \centering
        \includegraphics[width=0.8\textwidth]{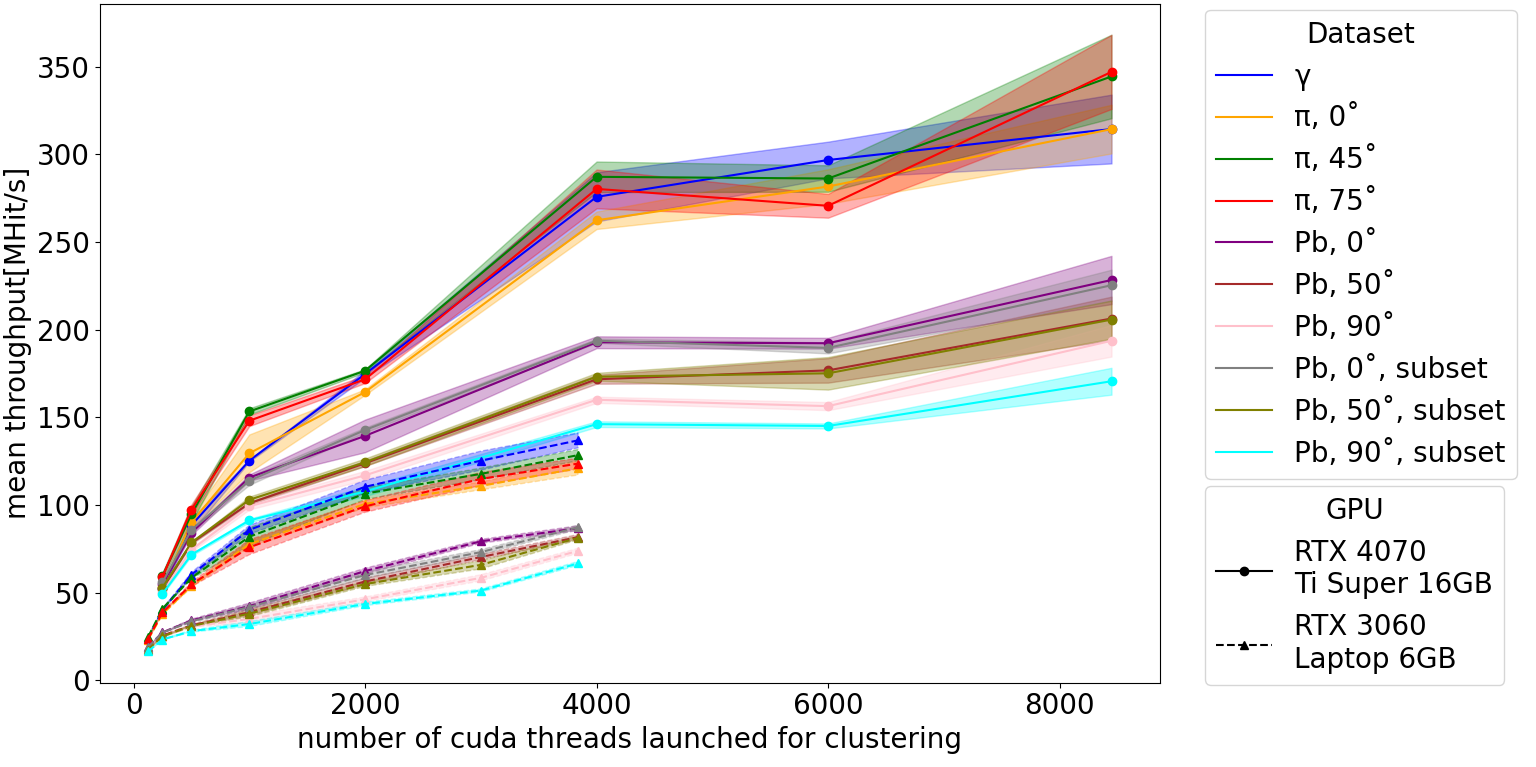}
        \caption{Dependence of parallel GPU clustering throughput on the number of launched cores. One of the tests was run on an Nvidia RTX 4070 Ti Super 16GB with PCIE 4.0$\times$16 with 8848 cores. The other test was run on an Nvidia RTX 3060 Laptop 6GB with PCIE 4.0$\times$4 with 3840 cores. The buffer size was set so that each thread, except the last one, processed a chunk with at least 10,000 hits. \label{fig:GPU benchmarking results}}

\end{figure}

The measured GPU clustering throughput is shown in Figure~\ref{fig:GPU benchmarking results}.
We used multiple optimization patterns to improve the clustering throughput. Pinned host memory~\cite{pinned_memory_copy_santos} enabled fast asynchronous memory transfers. Overlapping copy and compute using CUDA streams~\cite{cuda_streams_Li} hid the latency of data copying between the host and the device. Additionally, hit labels were consecutively updated in the low-latency shared memory and bulk-written to a global memory. To improve memory access locality, we also cached multiple hit data in shared memory. Lastly, the values in the auxiliary matrix were offset-encoded relative to the start of each chunk. Encoding each pixel using only two bytes, significantly reduced the memory usage per thread, thereby allowing for thread count saturation.

\section{Future work}

The research on clustering parallelization remains an active area with several unresolved challenges. In real-time clustering tasks performed on the CPU, there is a need to mitigate data loss that occurs due to high CPU utilization driven by the clustering process. 

When utilizing GPU-based clustering, a potential exists to further optimize parallelization by tailoring the clustering algorithm to specific data characteristics, such as handling small clusters. This specialization may enable a higher degree of parallelism. 
By relaxing the requirement that clusters remain temporally ordered, alternative strategies such as union by size or rank~\cite{union_find_tarjan} can be explored to enhance performance. Moreover, data permutation should be considered as a strategy to promote memory coalescence~\cite{memory_coalescence_fauzia} during reading operations. Although our initial experiments involving data permutation did not yield any performance improvements, further exploration may be needed.

Beyond clustering itself, other computational tasks, such as feature extraction, filtering, and particle identification using machine learning, can be offloaded to the GPU, potentially improving overall processing efficiency. An implementation of CPU-parallel clustering is available in Tracklab~\cite{tracklab_manek}. Additionally, a standalone implementation for both CPU and GPU clustering will be released in the near future; further information will be available on the website~\cite{celko_parallel_clustering_webpage}\cite{utef_software_webpage}.


\acknowledgments
The authors T.Č. and F.M. acknowledge funding from the Charles University Grant Agency (GAUK, project no. 142424). The authors B.B. and P.M. acknowledge funding from the Czech Science Foundation (GACR, reg. no. GM23-04869M). 





\bibliographystyle{JHEP}
 \bibliography{main}

\end{document}